# Nanoscale photonic network for solution searching and decision making problems

Makoto Naruse[†], Masashi Aono[††], and Song-Ju Kim[††]

[†] Photonic Network Research Institute, National Institute of Information and Communications Technology, Nukui-kita, Koganei, Tokyo 184-8795, Japan.

[††] RIKEN Advanced Science Institute, 2-1, Hirosawa, Wako, Saitama 351-0198, Japan

**SUMMARY** Nature-inspired devices and architectures are attracting considerable attention for various purposes, including the development of novel computing techniques based on spatiotemporal dynamics, exploiting stochastic processes for computing, and reducing energy dissipation. This paper demonstrates that networks of optical energy transfers between quantum nanostructures mediated by optical near-field interactions occurring at scales far below the wavelength of light could be utilized for solving a constraint satisfaction problem (CSP), the satisfiability problem (SAT), and a decision making problem. The optical energy transfer from smaller quantum dots to larger ones, which is a quantum stochastic process, depends on the existence of resonant energy levels between the quantum dots or a state-filling effect occurring at the larger quantum dots. Such a spatiotemporal mechanism yields different evolutions of energy transfer patterns in multi-quantum-dot systems. We numerically demonstrate that networks of optical energy transfers can be used for solution searching and decision making. We consider that such an approach paves the way to a novel physical informatics in which both coherent and dissipative processes are exploited, with low energy consumption.
*key words: nanophotonics, optical energy transfer, nature-inspired architecture, solution searching, decision making.*

## 1. Introduction

There is great demand for novel computing devices and architectures that can overcome the limitations of conventional technologies based solely on electron transfer, in terms of reducing power dissipation, solving computationally intractable problems, and so on [1]. Also, nature-inspired architectures are attracting significant attention from various research areas, such as brain-like computing and computational neurosciences [2], stochastic-based computing and noise-based logic [3], and spatiotemporal computation dynamics [4].

Among these research topics, Aono *et al*. demonstrated "amoeba-based computing" by utilizing the spatiotemporal oscillatory dynamics of the photoresponsive amoeboid organism Physarum combined with external optical control to solve a constraint satisfaction problem (CSP) [4] and the traveling salesman problem (TSP) [5]. Besides such experimental demonstrations, Leibtnitz *et al*. showed an algorithm for selecting the most suitable and robust network by utilizing fluctuations, inspired by biological experiments where the speed of fluorescence evolution of proteins in bacteria is observed to have a positive correlation with the phenotypic fluctuation of fluorescence over clone bacteria [6].

These demonstrations indicate that we can utilize the inherent spatial and temporal dynamics appearing in physical processes in nature for novel computing architectures and applications. Such arguments should also be applicable to nanometer-scale light–matter interactions. In fact, Naruse *et al*. demonstrated nanophotonic computing based on optical near-field processes at scales below the wavelength of light [7]. In particular, energy transfer between quantum nanostructures mediated by optical near-field interactions, detailed in Sec. 2 below, plays a crucial role. Optical near-field interactions, which are described by a Yukawa-type potential, have been used to realize energy transfer that involves conventionally dipole-forbidden energy levels. Its theoretical foundation has been explained by the dressed photon model [8], and the process has been experimentally demonstrated in various quantum nanostructures, such as InGaAs [9], ZnO [10], and CdSe [11]. In particular, Kawazoe *et al*. recently demonstrated room-temperature optical energy transfer using two-layer InGaAs quantum dots (QDs) [12]. In addition, the optical energy transfer has been shown to be $10^4$-times more energy efficient than that of the bit-flip energy required in conventional electrically wired devices [13].

This article demonstrates that a network of optical energy transfers between quantum dots mediated by optical near-field interactions can be utilized for solving the CSP, the satisfiability problem (SAT), and the multi-armed bandit problem (BP), which is a decision making problem. The optical energy transfer from smaller quantum dots to larger ones depends on the existence of resonant energy levels between the quantum dots or a state-filling effect occurring at the larger destination quantum dots. Also, as indicated by quantum master equations, the energy transfer process is fundamentally probabilistic. Such a spatiotemporal mechanism yields different evolutions of energy transfer patterns combined with certain control mechanisms, which we call bounceback control, similarly to the evolution of the shape of Physarum demonstrated by Aono *et al.* in Ref. [4]. At the same time, in contrast to biological organisms, optical energy transfer is implemented by highly controlled engineering means for designated structures, such as semiconductor quantum nanostructures fabricated by, for instance, molecular beam epitaxy [14] or DNA-based self-assembly [15]. The operating speed of such optical-near-field–mediated quantum dot systems, which is on the order of nanoseconds when radiative relaxation processes are involved, is significantly faster than those based on biological organisms, which is on the order of seconds or minutes [4,5]. The energy efficiency [13], as indicated already above, and the possibility of room-temperature operation [12] are also strong motivations behind the investigations described in this paper. In addition, we should emphasize that the concept and the principles discussed in this paper are fundamentally different from those of conventional optical computing or optical signal processing, which are limited by the properties of propagating light [16]. The concept and principles are also different from the quantum computing paradigm where a superposition of all possible states is exploited to lead to a correct solution [17]. The optical-near-field–mediated energy transfer is a coherent process, suggesting that an optical excitation could be transferred to all possible destination QDs via a resonant energy level, but such a coherent interaction between QDs results in a unidirectional energy transfer by an energy dissipation process occurring in the larger dot, as described in Sec. 2 below. Thus, our approach opens up the possibility of another computing paradigm where both coherent and dissipative processes are exploited.

This paper is organized as follows. Section 2 characterizes a nanoscale network of optical energy transfers via optical near-field interactions. Sections 3, 4, and 5 respectively demonstrate solving CSP, SAT, and decision making problems. Section 6 concludes the paper.

## 2. Nanoscale Network of Optical Energy Transfer

Here we assume two cubic quantum dots whose side-lengths are $a$ and $\sqrt{2}\,a$, which we call $QD_S$ and $QD_{L1}$, respectively, as shown in Fig. 1(a). There exists a resonance between the level of quantum number (1,1,1) in $QD_S$, denoted by S in Fig. 1(a), and that of quantum number (2,1,1) in $QD_{L1}$, denoted by $L_1^{(U)}$. Note that the (2,1,1)-level in $QD_{L1}$ is a dipole-forbidden energy level, meaning that propagating light cannot populate this level via optical excitations. However, optical near-fields allow this level to be populated thanks to the localized inhomogeneous fields in the vicinity of $QD_S$. Therefore, an exciton in the (1,1,1)-level in $QD_S$ could be transferred to the (2,1,1)-level in $QD_{L1}$. In $QD_{L1}$, due to the sublevel energy relaxation with a relaxation constant $\Gamma$, which is faster than the near-field interaction, the exciton relaxes to the (1,1,1)-level, denoted by $L_1^{(L)}$, from where it radiatively decays. As a result, we find unidirectional optical excitation transfer from $QD_S$ to $QD_{L1}$.

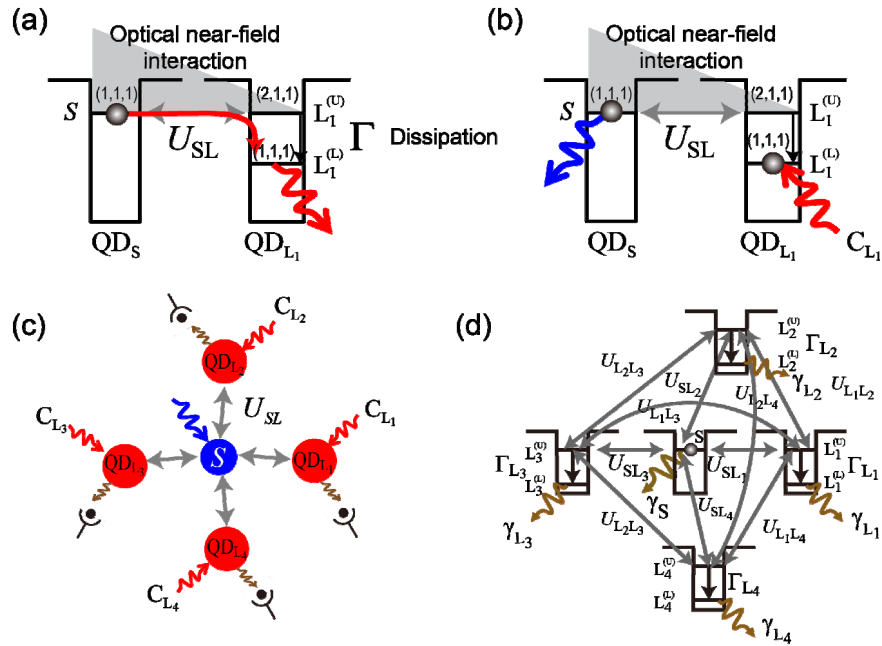

**Fig. 1** (a,b) Optical energy transfer between QDs mediated by near-field interactions. (c,d) Network of optical near-field interactions for solution searching.

When the lower energy level of the destination quantum dot is filled via another excitation (called "state filling"), an optical excitation occurring in a smaller QD ($QD_S$) cannot move to a larger one ($QD_{L1}$). This suggests two different patterns of optical energy transfer appear depending on the occupation of the destination quantum dot (Fig. 1(b)).

The key to achieving solution searching and decision making is to formulate a network of optical energy transfers. For instance, in the case of solving a CSP, shown in Sec. 3, we design an architecture where a smaller QD, labeled $QD_S$, is surrounded by four larger QDs, labeled $QD_{L1}$, $QD_{L2}$, $QD_{L3}$, and $QD_{L4}$, as indicated in Fig. 1(c). Fig. 1(d) shows representative parameterizations associated with the system. The (1,1,1)-level in $QD_S$ is denoted by S, and the (2,1,1)-level in $QD_{Li}$ is denoted by $L_i^{(U)}$. These levels are resonant with each other and are connected by inter-dot interactions denoted by $U_{SLi}$ ($i=1,...,4$). The lower level in $QD_{Li}$, namely the (1,1,1)-level, is denoted by $L_i^{(L)}$, which could be filled via the sublevel relaxation from $L_i^{(U)}$, denoted by $\Gamma_{Li}$. The radiations from the S and $L_i$ levels are respectively represented by the relaxation constants $\gamma_S$ and $\gamma_{Li}$. In the following description, we call the inverse of the relaxation constant the radiation lifetime. We also assume that the photon radiated from the lower level in $QD_{Li}$ can be separately captured by photodetectors. In addition, we assume control light beams, denoted by $C_{Li}$ in Fig. 1(c), that can induce a state filling effect at $L_i^{(L)}$. Summing up, Fig. 1(c) and (d) schematically represent the basic architecture of the system to be studied for solving a CSP, described in Sec. 3, and an SAT problem, described in Sec. 4.

First, we assume that the system initially has one exciton in S. From the initial state, through the inter-dot interactions $U_{SLi}$, the exciton in S can be transferred to $L_i^{(U)}$ ($i=1,...,4$). Correspondingly, we can derive quantum master equations in the density matrix formalism [8]. The Liouville equation for the system is then given by

$$\frac{d\rho(t)}{dt} = -\frac{i}{\hbar}[H_{int}, \rho(t)] - N_\Gamma \rho(t) - \rho(t) N_\Gamma, \quad (1)$$

where $\rho(t)$ is the density matrix with respect to the five energy levels, $H_{int}$ is the interaction Hamiltonian, and $N_\Gamma$ indicates relaxations. In the numerical calculation, we assume $U_{SLi}^{-1}=100$ ps, $\Gamma_i^{-1}=10$ ps, $\gamma_{Li}^{-1}=1$ ns, and $\gamma_S^{-1} \approx 2.92$ ns as a typical parameter set [18].

Based on the above modeling and parameterizations, we can calculate the populations involving $L_1^{(L)}$, $L_2^{(L)}$, $L_3^{(L)}$, and $L_4^{(L)}$, which are relevant to the radiation from the larger QDs. Also, when $QD_{Li}$ is subjected to state filling by control light $C_{Li}$, the energy transfer from $QD_S$ to $QD_{Li}$ behaves differently.

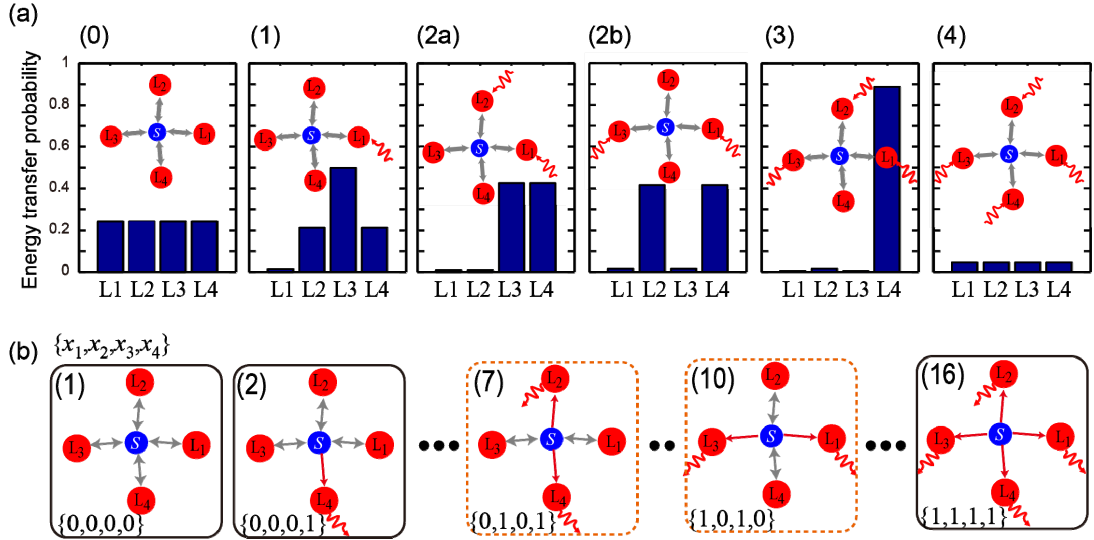

**Fig. 2** (a) Calculated energy transfer probabilities depending on the control light beams. (b) Schematic representation of possible states of the system. States (7) and (10) correspond to the correct solutions.

We assume that the probability of energy transfer to $QD_{Li}$ is correlated with the integral of the population of $L_i^{(L)}$, as summarized in Fig. 2(a). We should note that such integrals of the populations are indeed a figure-of-merit (FoM) indicating the trend of optical energy transfer from the smaller dot to the four larger ones. The law of conservation of probability does *not* hold; namely, the summation of the transition probabilities to $QD_{Li}$ is not unity. Instead, we see that the energy transfer to $QD_{Li}$ occurs if a random number generated uniformly between 0 and 1 is less than the transition probability to $QD_{Li}$ shown in Fig. 2(a); for example in the case of Fig. 2(a,(3)), the energy transfer to $QD_{L4}$ is highly likely, whereas the transfers to $QD_{L1}$, $QD_{L2}$, and $QD_{L3}$ are less likely.

The idea for problem solving is to control the optical energy transfer by controlling the destination QD using control light with a suitable mechanism, what we call *bounceback control*. The notion of bounceback, rather than feedback, implies that the system does *not* know the preferred status beforehand, in contrast to feedback control, which utilizes the difference between the present and the intended states.

## 3. Solving the Constraint Satisfaction Problem

We consider the following constraint satisfaction problem as an example regarding an array of $N$ binary-valued variables $x_i$ ($i=1,...,N$) [19]. The constraint is that $x_i=\text{NOR}(x_{i-1},x_{i+1})$ should be satisfied for all $i$. That is, variable $x_i$ should be consistent with a logical NOR operation of its two neighbors. For $i=0$ and $N$, the constraints are respectively given by $x_1=\text{NOR}(x_N,x_2)$ and $x_N=\text{NOR}(x_{N-1},x_1)$. We call this problem the "NOR problem" in this paper. Taking account of the nature of an individual NOR logic operation, one important inherent property is that, if $x_i=1$, then its two neighbors should both be zero, i.e., $x_{i-1}=x_{i+1}=0$. Now, we suppose that a photon radiated, or observed, from the energy level $L_i^{(L)}$ corresponds to a binary value $x_i=1$, whereas the absence of an observed photon means $x_i=0$. Therefore, $x_i=1$ should mean that optical energy transfer to both $L_{i-1}^{(L)}$ and $L_{i+1}^{(L)}$ is prohibited, so that $x_{i-1}=x_{i+1}=0$ is satisfied. Therefore, the bounceback mechanism is:

**[Bounceback rule for the NOR problem]** If $x_i=1$ at time $t$, then the control light beams $C_{i-1}$ and $C_{i+1}$ are turned on at time $t=t+1$.

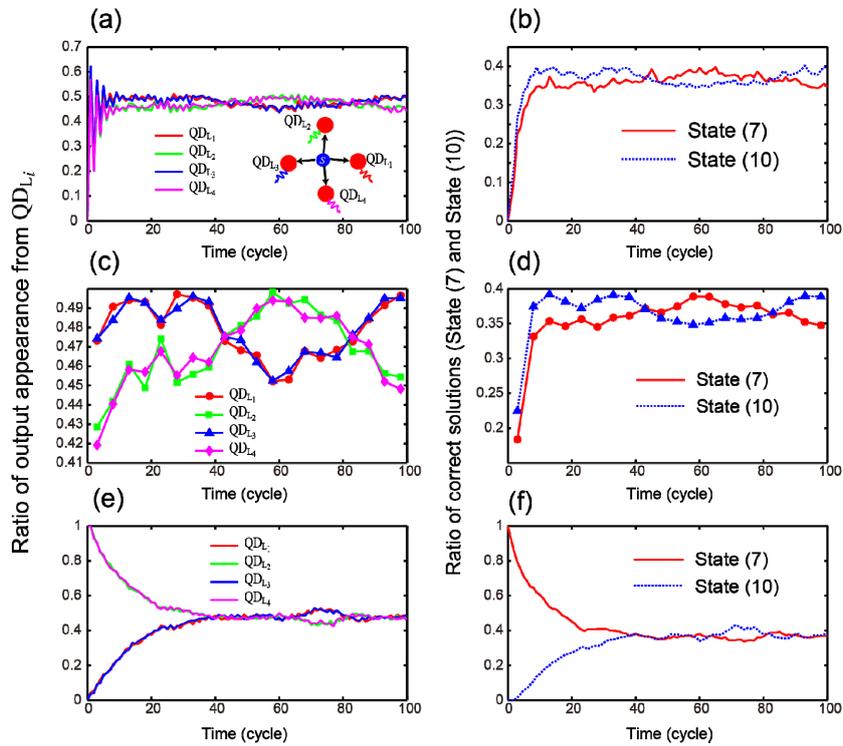

**Fig. 3** (a) The evolution of the ratio of the output appearance from $QD_{Li}$, and (b) the ratio of the states corresponding to correct solutions. (c,d) Time-averaged traces of (b) and (c), respectively. (e) The evolution of the ratio of the output appearance from $QD_{Li}$, and (f) the ratio of the states corresponding to correct solutions, with the initial state (7).

In the case of $N=4$, there are in total $2^4$ optical energy transfer patterns from the smaller dot to the larger ones. In this case, variables satisfying the constraints exist, and they are given by $\{x_1, x_2, x_3, x_4\}=\{0,1,0,1\}$ and $\{1,0,1,0\}$, which we call "correct solutions". Fig. 2(b) schematically represents some of the possible states, where States (7) and (10) respectively correspond to the correct solutions.

There are a few remarks that should be made regarding the NOR problem. One is about potential deadlock, analogous to Dijkstra's "dining philosophers problem", as already argued by Aono *et al*. in Ref. [4]. Starting with an initial state $x_i=0$ for all $i$, and assuming a situation where optical energy is transferred to all larger QDs, we observe photon radiation from all energy levels $L_i^{(L)}$, namely, $x_i=1$ for all $i$. Then, based on the bounceback mechanism shown above, all control light beams are turned on. If such a bounceback mechanism perfectly inhibits the optical energy transfer from the smaller QD to the large ones at the next step $t+1$, the variables then go to $x_i=0$ for all $i$. This leads to all control light beams being turned off at $t+2$. In this manner, all variables constantly repeat periodic switching between $x_i=0$ and $x_i=1$ in a synchronized manner. Consequently, the system can never reach the correct solutions. However, as indicated in Fig. 2(a), the probability of optical energy transfer to the larger dots is in fact not zero even when all larger QDs are illuminated by control light beams, as shown in Fig. 2(a,(4)). Also, even for a non-illuminated destination QD, the energy transfer probability may not be exactly unity. Such stochastic behavior of the optical energy transfer plays a key role in solving the NOR problem. This nature is similar to what was demonstrated in the amoeba-based computer [4], where fluctuations of chaotic oscillatory behavior involving spontaneous symmetry breaking in the amoeboid organism guarantees such a critical property.

The operating dynamics cause one pattern to change to another one every iteration cycle. Thanks to the stochastic nature, each trial could exhibit a different evolution of the energy transfer patterns. In particular, the transition probability, shown in Fig. 2(a), affects the behavior of the transitions. Therefore, we introduce a gain factor ($G$) to be multiplied by the energy transfer probability summarized in Fig. 2(a).

The curves in Fig. 3(a) represent the evolution of the output appearance from $QD_{Li}$, namely, the incidence ratio when $x_i=1$ among 1,000 trials evaluated at each cycle. The curves in Fig. 3(b) characterize the ratio of appearance of the states that correspond to the correct solutions: {0,1,0,1} (State (7)) and {1,0,1,0} (State (10)), respectively. When we closely examine the evolutions of $x_i$ in Fig. 3(a), we can see that the pair $x_1$ and $x_3$ exhibit similar behavior, as do the pair $x_2$ and $x_4$. Also, as the former pair exhibit larger values, the latter pair exhibit smaller values, and vice versa. This corresponds to the fact that correct solutions are likely to be induced as the number of iteration cycles increases.

Such a tendency is more clearly represented when we evaluate the time-averages of the characteristics in Fig. 3(a) and (b). Fig. 3(c) shows the evolutions of the ratio of the incidences when $x_i=1$, and Fig. 3(d) shows the ratios of State (7) and State (10) averaged over every 5 cycles. We can clearly observe a similar tendency to the one described above. Also, we should emphasize that, thanks to the probabilistic nature of the system, the states of correct solutions appear in an interchangeable manner. This is a clear indication of the fact that the probabilistic nature of the system autonomously seeks the solutions that satisfy the constraints of the NOR problem; the state-dependent probability of energy transfer plays a critical role in this. In other words, it should be emphasized that a non-local correlation is manifested in the evolution of $x_i$; for instance, when the system is in State (7), {0,1,0,1}, the probabilities of energy transfer to $QD_{L1}$ and $QD_{L3}$ are equally comparably low (due to state filling), whereas those to $QD_{L2}$ and $QD_{L4}$ are equally comparably high, indicating that the probability of energy transfer to an individual $QD_{Li}$ has inherent spatial patterns or non-local correlations. At the same time, the energy transfer to each $QD_{Li}$ is indeed probabilistic; therefore, the energy transfer probability to, for instance, $QD_{L1}$ is not zero even in State (7), and thus, the state could transition from State (7) to State (10), and vice versa. In fact, starting with the initial condition of State (7), the ratio of output appearance from $QD_{L1}$ and the ratio of the correct solutions evolve as shown in Fig. 3(e) and (f), where States (7) and (10) occur equally in the steady state at around 20 time cycles.

## 4. Solving the Satisfiability Problem (SAT)

SAT is the first problem proven to be nondeterministic polynomial time (NP)-complete, *i.e.*, the most difficult problem among those that belong to the complexity class NP [20]. Given a logical formula $\phi$, which consists of $N$ Boolean variables $x_i \in \{0(\text{false}), 1(\text{true})\}$ ($i \in I=\{1,2,...,N\}$), SAT is the problem of determining whether there exists at least one "satisfying" assignment of the truth values (0 or 1) to the variables represented by $x_i$ such that it makes the formula evaluate to true ($\phi=1$). Roughly speaking, $\phi$ represents a logical proposition, and the existence of a satisfying assignment verifies that the proposition is self-consistent. For example, the formula $\phi_{ex}=(x_1 \vee \neg x_2) \wedge (\neg x_2 \vee x_3 \vee \neg x_4) \wedge (x_1 \vee x_3) \wedge (x_2 \vee \neg x_3) \wedge (x_3 \vee \neg x_4) \wedge (\neg x_1 \vee x_4)$ has a unique solution $(x_1, x_2, x_3, x_4)=(1,1,1,1)$ that makes $\phi_{ex}=1$. This section describes a SAT problem solver inspired by the spatiotemporal dynamics of the network of optical energy transfers, what we call "NanoPS" [21].

SAT is called 3-SAT when $\phi$ consists of $M$ clauses that are connected by $\wedge$ (logical AND), and each clause connects at most three literals by $\vee$ (logical OR). Any SAT instance can be transformed into a 3-SAT instance, and 3-SAT is also NP-complete. A powerful SAT solver has great potential for a wide range of applications, such as artificial intelligence, information security, and bioinformatics, because the NP-completeness implies that all NP problems, including many practical real-world problems, can be transformed to the SAT problem [20].

In solving SAT by networks of optical excitation transfers, we assign two larger-sized quantum dots to a single variable $x_i$; namely, a $QD_L$ for representing $x_i=0$, and another $QD_L$ for $x_i=1$. Therefore, to solve $N$-variable 3-SAT, we use $2N$ $QD_L$s. $QD_{i,v}$ denotes the variable corresponding to $x_i=v$, where $v$ is either 0 or 1 and $i \in I=\{1,2,...,N\}$. When optical energy is transferred from $QD_S$ to $QD_{i,v}$ and radiation is subsequently observed at a time step $t$, we write this as $R_{i,v}(t)=1$, whereas $R_{i,v}(t)=0$ indicates that no radiation occurs. When state-filling stimulation is applied to $QD_{i,v}$, we denote this as $F_{i,v}(t)=1$, whereas $F_{i,v}(t)=0$ denotes no state-filling. As discussed in Sec. 2 and 3, radiation from $QD_{i,v}$ depends stochastically on the energy transfer probability denoted by $p_{i,v}$ as follows:

$$R_{i,v}(t) = \begin{cases} 1 & \text{with probability } p_{i,v} \text{ if } F_{i,v}(t) = 1 \\ 1 & \text{with probability } 1 - p_{i,v} \text{ if } F_{i,v}(t) = 0 \\ 0 & \text{otherwise.} \end{cases} \quad (2)$$

Each radiation event $R_{i,v}(t)$ is accumulated by a newly introduced variable $X_{i,v}(t) \in \{-1,0,1\}$ as follows:

$$X_{i,v}(t+1) = \begin{cases} X_{i,v}(t)+1 & \text{if } R_{i,v}(t) = 1 \text{ and } X_{i,v}(t) < 1 \\ X_{i,v}(t)-1 & \text{if } R_{i,v}(t) = 0 \text{ and } X_{i,v}(t) > -1 \\ X_{i,v}(t) & \text{otherwise.} \end{cases} \quad (3)$$

The above dynamics can be implemented either in the form of combinations of QDs or external electrical circuits. At each step $t$, the variable $X_{i,v}(t)$ yields the estimated variables $x_i$ as follows:

$$x_i(t) = \begin{cases} 0 & \text{if } X_{i,0}(t) = 1 \text{ and } X_{i,1}(t) \leq 0 \\ 1 & \text{if } X_{i,1}(t) \leq 0 \text{ and } X_{i,v}(t) = 1 \\ x_i(t-1) & \text{otherwise.} \end{cases} \quad (4)$$

The state-filling stimulations $F_{i,v}$ are updated synchronously according to the following dynamics:

$$F_{i,v}(t+1) = \begin{cases} 1 & \text{if } \exists (P,Q) \in B \text{ s.t.} \\ & \forall (j,u) \in P, X_{j,u}(t) = 1 \text{ and } (i,v) \in Q \\ 0 & \text{otherwise} \end{cases} \quad (5)$$

where $B$ is a set of bounceback rules to be defined shortly. Each element $(P,Q)$ in $B$ implies the following statement: if all the $X_{j,u}$s specified by $P$ are positive at step $t$, then stimulate all QD$_{i,v}$s specified by $Q$ to inhibit their radiation at step $t+1$.

To understand the meaning of the bounceback rules, let us consider the example formula $\phi_{ex}$. To satisfy the formula $\phi_{ex}=1$, every clause in $\phi_{ex}$ should be true. For example, suppose the system tries to assign $x_1=0$, i.e., $X_{1,0}(t)=1$, as indicated by the black broken circle in Fig. 4. Now let us focus on the first clause $(x_1 \vee \neg x_2)$ of $\phi_{ex}$. To make this clause true, if $x_1=0$ then $x_2$ should not be 1. Therefore, we apply state-filling stimulation $F_{2,1}(t+1)=1$ to inhibit radiation $R_{2,1}(t+1)$ from QD$_{2,1}$, as indicated by the gray broken circle in Fig. 4. At the same time, $x_3$ in the third clause $(x_1 \vee x_3)$ should not be 0, and so we apply $F_{3,0}(t+1)=1$ (the gray dotted circle). In addition, we must apply $F_{1,1}(t+1)=1$ (the solid circle) because $x_1=0$ necessitates that $x_1$ should not be 1. Likewise, the set of all bounceback rules $B$ is determined by scanning all clauses in $\phi_{ex}$.

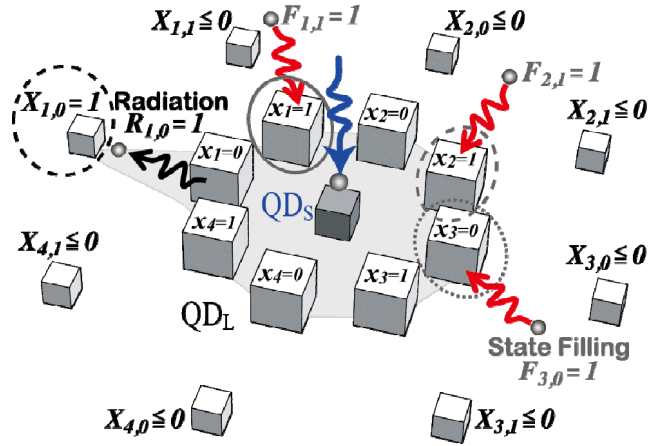

**Fig. 4** Schematic representation of the QD formation and the bounceback control for the solution search of a four-variable SAT problem. The bounceback control applies state-filling stimulations $F_{1,1}(t+1)=F_{2,1}(t+1)=F_{3,0}(t+1)=1$ if $X_{1,0}(t)=1$.

The elements of the bounceback rule $B$ formally consist of three parts: $B=\text{INTRA} \cup \text{INTER} \cup \text{CONTRA}$.
INTRA forbids each variable $i$ from taking two values 0 and 1 simultaneously:

$$\text{INTRA} = \{\{(i,v)\},\{(i,1-v)\}) \mid i \in I \wedge v \in \{0,1\}\}. \quad (6)$$

Each clause $c=(x_j^* \vee x_k^* \vee x_l^*)$ has its literals $x_i^*$ mapped to $i^*=i$ if $x_i^*=x_i$ and to $-i$ otherwise, and the formula $\phi$ is expressed equivalently by a set $\Phi$, which includes all the clauses as its elements. For example, the example formula $\phi_{ex}$ is transformed into $\Phi_{ex}=\{\{1,-2\},\{-2,3,-4\},\{1,3\},\{2,-3\},\{3,-4\},\{-1,4\}\}$. For each $C$ in $\Phi$ and each variable $i$ in $C$, INTER blocks the radiation that makes $C$ false [either $R_{i,0}(t+1)$ or $R_{i,1}(t+1)$]:

$$\text{INTER} = \{(P,\{(i,0)\}) \mid i \in C\} \cup \{(P,\{(i,1)\}) \mid -i \in C\} \quad (7)$$

where $P=\{(j,0) \mid j \in C \wedge j \neq i\} \cup \{(j,1) \mid \neg j \in C \wedge j \neq i\}$. Some rules in INTER may imply that neither 0 nor 1 can be assigned to a variable. To avoid this contradiction, for each variable $i$, we build CONTRA by checking all the relevant rules in INTER:

$$\text{CONTRA} = \{(P \cup P', P \cup P') \mid i \in I \wedge (P, \{(i,0)\}) \in \text{INTER} \wedge (P', \{(i,1)\}) \in \text{INTER}\}. \tag{8}$$

Before we start solving the given problem, $B$ is obtained in polynomial time $O(N \bullet M)$ by generating all the bounceback rules in INTRA, INTER, and CONTRA based on the above procedures.

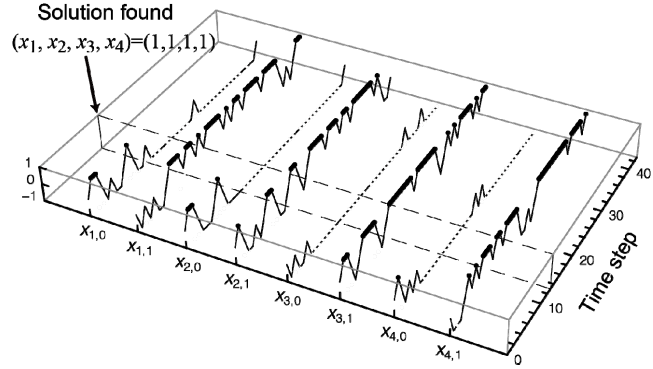

**Fig. 5** Simulated time evolution of $X_{i,v}(t)$ in the SAT solver using a network of optical near-fields. The system found the solution $(x_1, x_2, x_3, x_4)=(1,1,1,1)$ at $t=13$.

The calculation starts from the condition that $X_{i,v}(0)=R_{i,v}(0)=F_{i,v}(0)=0$ for all $(i,v)$, and the time evolution of the system is simulated by calculating the above equations iteratively. Fig. 5 shows that the system successfully found the solution of the example formula $\phi_{ex}$ at step $t=13$.

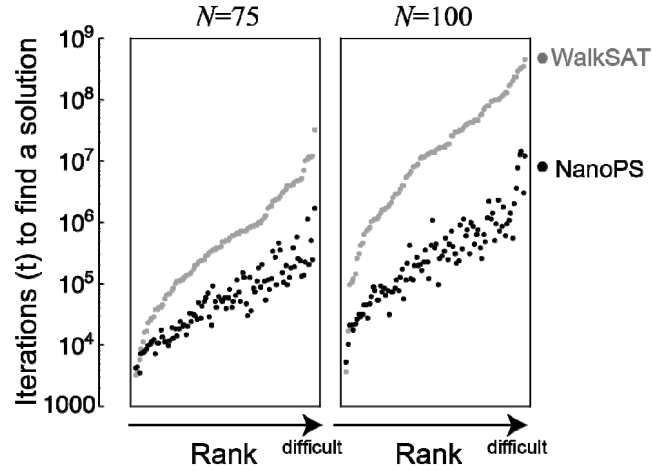

**Fig. 6** Performance comparison of NanoPS and WalkSAT. For each algorithm and each $N$, 100 instances were evaluated. The instances were sorted from easiest to the most difficult in ascending order of the average number of iterations that WalkSAT required to find a solution.

The performance of the proposed NanoPS was compared with that of the best-known search algorithm, WalkSAT [22]. In WalkSAT, an assignment $x(0)=(x_1(0), x_2(0),...,x_N(0))$ is initially randomly chosen. At each time step $t$, by checking whether each clause is satisfied by the current assignment $x(t)$, WalkSAT randomly chooses one of the unsatisfied clauses and satisfies it by flipping one of its variables chosen at random. This routine is iterated until a satisfying assignment is obtained. Schöning estimated the average number of iterations that WalkSAT required for finding a solution to a 3-SAT problem as the exponential function $(4/3)^N \text{poly}(N)$ [23].

We used benchmark problem instances provided by SATLIB online [24], which were the most difficult 3-SAT instances obtained by randomly generating three-literal conjunctive normal form formulae, where the difficulty can be maximized by setting the ratio between the number of variables $N$ and the number of clauses $M$ at the phase transition region around $M/N=4.26$ [25,26]. We chose 100 instances from each of the test sets uf75-325 and uf100-430, which took [$N=75$, $M=325$] and [$N=100$, $M=430$] formulae from the most difficult region where $M/N$ is about 4.333 and 4.3, respectively.

For each instance, we conducted 500 trials consisting of Monte Carlo simulations to obtain the average number of iterations (time steps $t$) required to find a solution. As shown in Fig. 6, NanoPS, which is a nanophotonic network, found a solution after a much smaller number of iterations than WalkSAT. Also, the advantage of NanoPS over WalkSAT increased as the number of variables $N$ increased.

## 5. Solving the Decision Making Problem

Consider a number of slot machines, each of which rewards the player with a coin at a certain probability $P_k$ ($k \in \{1,2,...,N\}$) when played. To maximize the total amount of reward, it is necessary to make a quick and accurate judgment of which machine has the highest probability of giving a reward. To accomplish this, the player should gather information about many machines; however, in this process, the player should not fail to exploit the reward from the known best machine. These requirements are not easily met simultaneously because there is a trade-off between "exploration" and "exploitation", referred to as the "exploration–exploitation dilemma". Such a problem is called the multi-armed bandit problem (BP).

BP was originally described by Robbins [27], although the same problem in essence was also studied by Thompson [28]. However, the optimal strategy is known only for a limited class of problems in which the reward distributions are assumed to be known to the players [29,30]. There are a number of important practical applications of BP, such as Monte Carlo tree searches [31].

Biological organisms commonly encounter the "exploration–exploitation dilemma" in surviving uncertain environments. Inspired by the amoeba's shape-changing process under dynamic light stimuli, Kim *et al.* proposed an algorithm for BP called the "tug-of-war model" (TOW) [32].

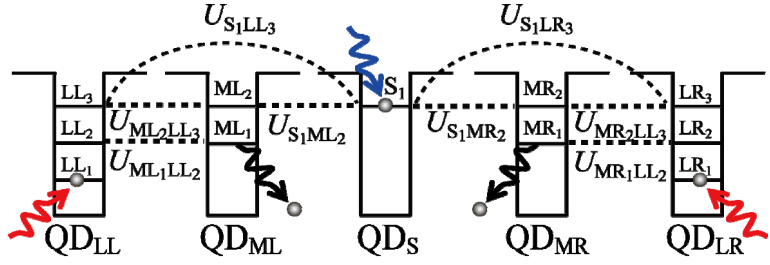

**Fig. 7** Nanophotonic decision maker (NanoDM) composed of five QDs mediated by inter-dot optical near-field interactions.

TOW is a dynamical system model of an amoeba-like body, which maintains a constant intracellular resource volume while collecting environmental information by concurrently expanding and shrinking its branches. The conservation law entails a "nonlocal correlation" among the branches; that is, the volume increment in one branch is immediately compensated by volume decrement(s) in the other branch(es). This nonlocal correlation was shown to enhance the performance in solving BP [32].

Here we show that a network of optical energy transfers among quantum dots can implement a variant of TOW, which we call a "nanophotonic decision maker" (NanoDM) [33]. Although we demonstrate only the two-armed case, NanoDM can be easily extended to $N$-armed ($N>2$) cases.

We use three types of cubic QDs with side lengths $a$, $\sqrt{2}\,a$, and $2a$, which are represented by $QD_S$, $QD_M$, and $QD_L$, respectively. We assume that five QDs are one-dimensionally arranged in the order $QD_L$-$QD_M$-$QD_S$-$QD_M$-$QD_L$. When an optical excitation is generated in $QD_S$, it is transferred to the lowest energy levels in both $QD_L$s through the inter-dot optical near-field interaction network; thus we observe negligible radiation from the $QD_M$s. However, when the lowest energy levels of the $QD_L$s are populated by control light, which induce state-filling effects, an exciton at $QD_S$ is more likely to be radiated from $QD_M$.

We consider the radiation from the $QD_M$s, namely, the left one ($QD_{ML}$) or right one ($QD_{MR}$), as the decision of selecting slot machine A or B, respectively. The intensity of the control light to induce state-filling at the left and right $QD_L$s is respectively modulated on the basis of the resultant rewards obtained from the chosen slot machine. Similar to the demonstrations shown in Sec. 3 and 4, the fundamentally probabilistic attributes of optical energy transfer in a multi-quantum-dot system are exploited for the decision making.

In NanoDM, there are in total 11 energy levels, as schematically shown in Fig. 7. The energy levels are networked either by optical near-field interactions or sublevel energy dissipation. The populations concerning the radiation from the $QD_M$s are calculated based on a density matrix formalism taking account of the external control light radiating the $QD_L$s.

We adopt an intensity adjuster (IA) to modulate the intensity of incident light to the $QD_L$s, as shown at the bottom of Fig. 8. Also, we consider that the effects of state filling can be equivalently represented by the value of sublevel relaxation parameters, which has been validated in Ref. [19].

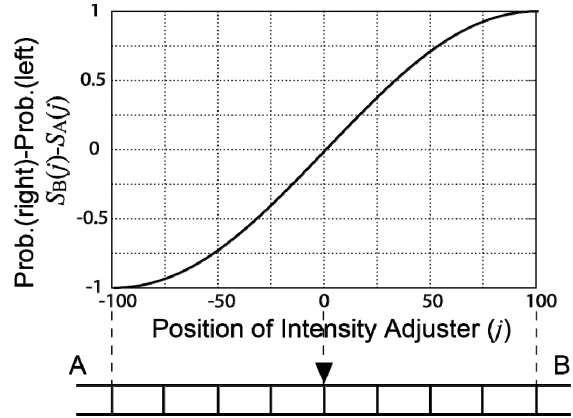

**Fig. 8** Intensity adjuster (IA) and the difference between radiation probabilities from $QD_{MR}$ and $QD_{ML}$.

The initial position of the IA is zero. In this case, the same intensity of light is applied to both the energy levels $LL_1$ and $LR_1$ shown in Fig. 7. If we move the IA to the right, the intensity at the right increases and that at the left decreases. In contrast, if we move the IA to the left, the intensity at the left increases and that at the right decreases. This situation can be described by the following relaxation rate parameters as functions of the IA position $j$: $\Gamma_{LR2}=1/100-j/10000+1/100000$ and $\Gamma_{LL2}=1/100+j/10000+1/100000$. The radiation probabilities from $ML_1$ and $MR_1$, which are respectively denoted as $S_A(j)$ and $S_B(j)$, are derived by solving the master equation. The difference between radiation probabilities, $S_B(j)-S_A(j)$, is shown by the solid line in Fig. 8.

The dynamics of the IA are defined as follows:
1. Set the IA position $j$ to 0.
2. Select machine A or B based on $S_A(j)$ and $S_B(j)$.
3. Play the selected machine.
4. If a coin is dispensed, then move the IA in the direction of the selected machine, that is, $j=j-D$ for A and $j=j+D$ for B, where $D$ is the amount of the increment.
5. If no coin is dispensed, then move the IA in the opposite direction from the selected machine, that is, $j=j+D$ for A and $j=j-D$ for B.
6. Go back to step 2.

In this way, NanoDM selects A or B, and the IA moves to the right or left according to the reward.

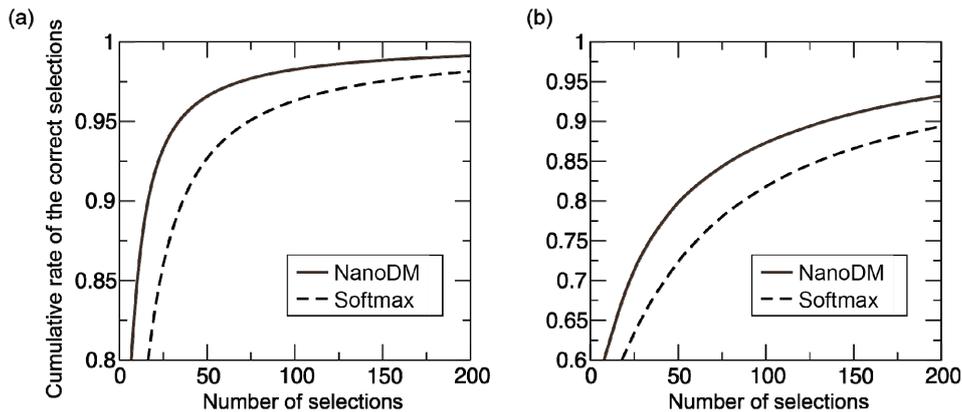

**Fig. 9** Performance comparison between NanoDM and the Softmax algorithm. (a) $P_A=0.2$, $P_B=0.8$; (b) $P_A=0.4$, $P_B=0.6$.

We compared the performance of NanoDM with that of the Softmax algorithm, which is known to be the best-fitting algorithm for human decision-making behavior in the BP [34]. Figs. 9(a) and (b) demonstrate the efficiency (cumulative rate of correct selections) for NanoDM (solid line) and Softmax with the optimized parameter (broken line) in the case where the reward probabilities of the slot machines are (a) $P_A=0.2$ and $P_B=0.8$ and (b) $P_A=0.4$ and $P_B=0.6$. In these cases, the correct selection is "B" because $P_B$ is greater than $P_A$. These cumulative rates of correct selections are average values for each 1,000 samples. Hence, each value corresponds to the average number of coins acquired from the slot machines. Even with a nonoptimized parameter $D$, the performance of NanoDM was higher than that of Softmax with its optimized parameter in a wide parameter range of $D$=10 to 100, although we show only the $D$=50 case in Figs. 9(a) and (b).

One remark is that in this study we dealt with restricted problems, namely, $P_A+P_B$=1. General problems can, however, also be solved by an extended NanoDM, although the IA dynamics become slightly complicated.

## 6. Conclusion

In summary, we have demonstrated that a nanoscale network of optical energy transfers between quantum nanostructures mediated by optical near-field interactions occurring at scales far below the wavelength of light has the potential to solve solution searching and decision making problems. More specifically, we demonstrated solving a constraint satisfaction problem, a satisfiability problem, and a multi-armed bandit problem. The key is that nanostructured matter in the form of quantum dots are networked via optical near-fields; optical energy transfer from smaller quantum dots to larger ones, which is a quantum stochastic process, depends on the existence of resonant energy levels between the quantum dots or a state-filling effect occurring at the destination quantum dots. We exploit these unique spatiotemporal mechanisms in optical energy transfer to solve solution searching and decision making problems.

As indicated in the introduction, the concept and the principles demonstrated in this paper are based on both coherent and dissipative processes on the nanoscale, which is not the case with conventional optical, electrical, and quantum computing paradigms. The inherently non-local nature is also a unique attribute provided by the optical-near-field–mediated optical energy transfer network. This work shown in this paper paves the way to applying nanometer-scale photonic networks to solving computationally demanding applications and suggests a new computing paradigm.

## Acknowledgments

The authors would like to thank many collaborators involved in this work, in particular Drs. M. Ohtsu, T. Kawazoe, W. Nomura, H. Hori, and M. Hara.